\documentclass{nature1}

\newcommand {\gtrsim} {\ {\raise-.5ex\hbox{$\buildrel>\over\sim$}}\ }
\newcommand {\ltrsim} {\ {\raise-.5ex\hbox{$\buildrel<\over\sim$}}\ }
\usepackage{graphicx}

\def\apj{{\it Astrophys. J.}}

\def\apjsup{{\it Astrophys. J. Suppl.}}

\def\aj{{\it Astron. J.}}

\def\mnras{{\it Mon. Not. R. Astron. Soc.}}

\def\aa{{\it Astron. \& Astrophys.}}

\def\aasup{{\it Astron. \& Astrophys. Suppl.}}

\def\science{{\it Science}}

\def\araa{{\it Ann. Rev. Astron. Astrophys.}}

\def\pasp{{\it Publ. Astron. Soc. Pacific}}

\def\oiil{[OII]}
\def\oiiil{[OIII]}

\title{Lyman continuum leaking from the compact 
star-forming dwarf galaxy J0925+1403}

\author{Y. I. Izotov$^1$, I. Orlitov\'a$^2$, 
D. Schaerer$^{3,4}$, T. X. Thuan$^5$, A. Verhamme$^3$, N. G. Guseva$^1$ \& G. Worseck$^6$}

\begin{document}
\spacing{1}

\maketitle

\begin{affiliations}
 \item {\rm \ Main Astronomical Observatory, National Academy of Sciences of Ukraine, 
27 Zabolotnoho str., Kyiv 03680, Ukraine}
 \item {\rm \ Astronomical Institute, Czech Academy of Sciences, Bo\v cn{\'\i} II 1401, 141 00, Prague, Czech Republic}
 \item {\rm \ Observatoire de Gen\`eve, Universit\'e de Gen\`eve, 
51 Ch. des Maillettes, 1290, Versoix, Switzerland}
\item {\rm \ CNRS, IRAP, 14 Avenue E. Belin, 31400 Toulouse, France}
 \item {\rm \ Astronomy Department, University of Virginia, PO Box 400325, Charlottesville, VA 22904, USA}
 \item{\rm \ Max-Planck-Institut f\"ur Astronomie, K\"onigstuhl 17, 69117 Heidelberg, Germany}
\end{affiliations}

\begin{abstract}
One of the key questions in observational cosmology is the
identification of
the sources responsible for ionisation of the Universe after the cosmic
Dark Ages,
when the baryonic matter was neutral. The currently
identified distant galaxies are insufficient to fully reionise
the Universe by redshift $z \sim 6$\cite{S01,Iw09,R13},
but low-mass star-forming galaxies are thought to be responsible for the
bulk of the ionising radiation\cite{M13,Y11,WC09}.
Since direct observations at high redshift are difficult for a variety of reasons,
one solution is to identify local proxies of this galaxy population.
However, starburst galaxies at low redshifts are generally opaque to
their ionising radiation\cite{L95,D01,Gr09}. This radiation with
small escape fractions of $\sim$1-3 \% is directly detected only in three 
low-redshift galaxies\cite{L13,B14}.
Here we present far-ultraviolet
observations of a nearby low-mass star-forming galaxy, J0925+1403, selected
for its compactness and high excitation.
The galaxy is leaking ionising radiation, with an escape fraction of $\sim$ 8\%.
The total number of photons emitted during the starburst phase is sufficient 
to ionize 
intergalactic medium material, which is about 40 times more massive than
the stellar mass of the galaxy.
\end{abstract}

So-called ``Green Peas'' (GP), low-mass  compact galaxies with very
active star formation\cite{Ca09,I11,JO13,S15},
may be promising candidates for escaping ionising radiation.
The GP galaxy J0925$+$1403 was
selected from the Sloan Digital Sky Survey (SDSS) according
to the following properties (Methods section):
1) a compact structure;
2) the presence of emission lines with high equivalent widths in its SDSS spectrum, 
suggesting active ongoing star formation and
numerous hot O stars producing ionising Lyman continuum (LyC) radiation;
3) sufficiently bright in the far-ultraviolet (FUV) with a magnitude of 20.7 mag 
and redshifted enough ($z$ = 0.301) to allow direct LyC observations
with the Cosmic Origins Spectrograph (COS) onboard the Hubble Space
Telescope (HST); and
4) a high O$_{32}$ = [OIII]$\lambda$5007/[OII]$\lambda$3727
flux ratio of 5 (see Fig.\ 1), which may indicate the presence of density-bounded 
HII regions\cite{NO14}, i.e.\  escaping LyC radiation.

We first derive some general properties of the galaxy, using the emission-line fluxes 
measured from the SDSS optical spectrum. After correction for 
the Milky Way extinction of $A_{V, {\rm MW}}$ = 0.084 mag, we obtain an internal 
extinction $A_{V, {\rm int}}$ = 0.36 mag, 
and a low oxygen abundance  12~+~logO/H = 7.91$\pm$0.03, or less than 0.2 solar. 
The details of these determinations are given in the Methods section.
Everywhere in the paper the errors are 1$\sigma$ errors.

The same SDSS spectrum is used to fit a spectral energy distribution (SED) to
derive the galaxy's global parameters, including the stellar mass and the age of 
the present burst of star formation (see Methods section). 
We obtain a starburst age of 2.6$\pm$0.2 Myr,  a young stellar 
mass of (2.4$\pm$0.3)$\times$10$^8$~$M_\odot$, and a total galaxy stellar mass of 
(8.2$\pm$0.7)$\times$10$^8$~$M_\odot$.
The star-formation rate is 52.2~$M_\odot$~yr$^{-1}$, as determined from the 
extinction-corrected H$\beta$ line flux.
With its low mass, low metallicity, low 
extinction, compact morphology, and high star-formation rate, J0925$+$1403 
shares many of the properties of high-redshift Lyman Alpha Emitters.

GPs with  O$_{32}$ $\geq5$ have been observed before by HST\cite{JO14,H15}, but 
their low redshifts $z<0.3$ were not optimal for LyC observations.
The HST/COS observations of J0925$+$1403 were obtained on 
28 March, 2015 (program GO13744, PI: T. X. Thuan). 
The near-ultraviolet (NUV) acquisition image shows the galaxy to have a very compact 
structure, with a half-light angular diameter of $\sim $0.2$''$, much smaller than the 
spectroscopic aperture of 2.5$''$ (Fig. 2). 
This angular diameter corresponds to a linear diameter of $\sim$ 1 kpc at the 
angular diameter distance of 930 Mpc, derived from the redshift $z$ = 0.301, adopting the
Planck mission cosmological parameters $H_0$ = 67.1 km s$^{-1}$Mpc$^{-1}$,
$\Omega_\lambda$ = 0.682 and $\Omega_m$ = 0.318\cite{P14}.

Spectra of J0925$+$1403 were obtained with two gratings. The low-resolution 
G140L grating ($<$900 -- 2385 \AA) was used to obtain the spectrum, which 
includes the redshifted LyC emission, with an exposure time of 5649 s. The 
medium-resolution G160M 
grating (1410 -- 1796 \AA) was used to obtain the spectrum, which includes the redshifted 
Ly$\alpha$ $\lambda$1216 \AA\ line, with the exposure time of 2978 s.
The observations with the G160M and G140L gratings were reduced with the standard 
pipeline and custom software, respectively. The custom software gives more accurate 
results, as it is specifically designed for faint HST/COS targets (see Methods section).

A strong Ly$\alpha$ $\lambda$1216\,\AA\ emission-line is detected in the 
medium-resolution spectrum. Its profile (Fig.~3) shows  two 
peaks on both sides of the line center (dashed vertical line).    
According to radiative transfer models, the separation between the
 Ly$\alpha$ line peaks increases 
with increasing optical depth and thus with increasing neutral 
hydrogen column density $N$(HI)\cite{V15}. 
In the case of J0925$+$1403, the separation of $\sim$ 300 km s$^{-1}$ 
corresponds to a low column density 
($\log N{\rm (HI)} \leq 10^{19}$ cm$^{-2}$), allowing the escape of a considerable
fraction of the Ly$\alpha$ emission.
Correcting for the Milky Way and galaxy internal reddening
we obtain a Ly$\alpha$ flux density of 8.2$\times$10$^{-14}$ erg s$^{-1}$cm$^{-2}$. 
Comparing the extinction-corrected Ly$\alpha$/H$\beta$ flux
ratio of 16.7$\pm$1.0 and case B flux ratio of 23.3\cite{HS87}, we find that the 
Ly$\alpha$ escape fraction is $\sim$70\%, among the highest known so far for GP 
galaxies\cite{H15}, and consistent with a low HI column density.

The short-wavelength part of the J0925$+$1403 spectrum,
obtained with the low-resolution grating G140L, is shown in Fig. 4a by the grey solid line. 
The modelled UV SED of the young cluster 
with the age and extinction parameters obtained before  
from SED fitting in the optical range (see Method section)
is shown by the black solid line. We adopt the reddening curve\cite{C89}, 
corresponding to $R_{V, {\rm MW}}$ = $A_V$/$E_{B-V}$ = 3.1 for the Milky Way 
and  to $R_{V, {\rm int}}$ = 2.4 for J0925$+$1403, 
except for $\lambda\leq$ 1250\,\AA, where the reddening curve\cite{M90} is used.
The corresponding intrinsic
spectral energy distribution is shown in Fig. 4a by the black dash-dotted line. 
The flux density of the intrinsic Lyman continuum
is determined primarily by the extinction-corrected flux density of the H$\beta$ emission line
and the starburst age. The starburst age is derived from the condition that the observed
equivalent width of the H$\beta$ emission line is equal to the modelled value, which
depends on the extinction-corrected flux of the continuum near H$\beta$ and the
intrinsic LyC flux. The intrinsic LyC  is fairly insensitive to the adopted stellar evolution 
models, stellar atmosphere models and initial mass function (see Method section).
It is seen that the reddened SED reproduces very well the observed spectrum for rest-frame
wavelengths $>$ 912\AA.

Fig. 4b shows a blow-up of the LyC spectral region.
The important feature to note is that the Lyman continuum flux density for 
$\lambda$ $<$ 912\AA\ is not zero, but positive with a value equal to 
(2.35$\pm$0.20)$\times$10$^{-17}$ erg s$^{-1}$ cm$^{-2}$ \AA$^{-1}$, when
averaged over the 861 -- 907 \AA\ rest-frame spectral range. 
It is detected at the 11.6$\sigma$ level and is indicated 
by a dotted horizontal line and a filled circle in Fig. 4b. 
This observed LyC should be corrected
for the Milky Way extinction before the determination of the LyC escape
fraction. The extinction-corrected average LyC flux density of 
(3.43$\pm$0.29)$\times$10$^{-17}$ erg s$^{-1}$ cm$^{-2}$ \AA$^{-1}$ is shown
by the thick solid horizontal line (Fig. 4b).
Comparing this value to the intrinsic continuum flux density 
of 4.4$\times$10$^{-16}$ erg s$^{-1}$ cm$^{-2}$ \AA$^{-1}$ beyond the Lyman
limit for $\lambda<$912 \AA, we obtain an absolute LyC escape fraction 
$f_{\rm esc}$ = 7.8\%$\pm$1.1\% in J0925$+$1403, where the error is determined
by the observed LyC flux density error and uncertainties in the modelled intrinsic LyC flux
density. This value is several times higher than $f_{\rm esc}$ of the other three low-redshift 
galaxies with known LyC leakage. J0925+1403 also has the highest \oiiil/\oiil\ flux ratio, 
the lowest metallicity and the lowest stellar mass. 
Thus we conclude that compact low-mass star-forming galaxies with high \oiiil/\oiil\ ratios 
may lose a considerable fraction of their LyC emission to the IGM.

The above determination of $f_{\rm esc}$ holds for UV-emitting star-forming regions without 
dust-obscured star formation, which is invisible in the UV and/or optical ranges. 
The sky region containing the galaxy J0925$+$1403 has been observed in the mid-infrared range
by the Wide-field Infrared Survey Explorer (WISE). However, data for this galaxy are not present in the AllWISE Source 
Catalog\cite{WISE13}. There are also
no data in the radio range. Optical, near- and mid-infrared observations of other
low-metallicity star-forming galaxies with similar properties suggest that they are 
relatively transparent\cite{IT11}. Negligible dust-obscured
star formation is also implied by the observed thermal free-free cm radio emission in dwarf 
galaxies with optical and infrared observations as it is consistent with the value derived 
from the flux density of the H$\beta$ emission line\cite{I14}. Presumably, the same 
conclusion holds for J0925$+$1403.

The number of ionising photons escaping the galaxy 
is $Q_{\rm H}$ = 3.86$\times$10$^{53}$~s$^{-1}$ if $f_{\rm esc}$ = 7.8~\% 
(Methods section), corresponding to a total number of ionising photons of 
3.6$\times$10$^{67}$
emitted during a starburst with a 3 Myr duration. This total number of photons is
sufficient to ionise the low-density IGM gas with a mass of 
$\sim$4$\times$10$^{10}$ $M_\odot$, or about
40 times higher than the stellar mass of the galaxy, assuming one photon
suffices to ionise one hydrogen atom.
Here we adopt the luminosity distance of 1620 Mpc.
Finally, we note that our galaxy leaks a large number of ionising photons per UV luminosity, 
$Q_{\rm H}/L_{1500} \approx 10^{25}$ photon $s^{-1} / ({\rm ergs\ s^{-1} Hz^{-1}})$,  
approximately three times more, than (optimistic) assumptions\cite{R13} 
used at high redshift for $f_{\rm esc}=0.2$,
which is primarily due to the young age of the UV-dominant stellar population in 
J0925$+$1403.

\subsection{Online Content} Methods, along with any additional Extended 
Data display items and Source Data, are available in the online version of 
the paper; references unique to these sections appear only in the online paper.



\begin{addendum}
 \item Based on observations made with the NASA/ESA Hubble Space Telescope, 
obtained from the data archive at the Space Telescope Science Institute. 
STScI is operated by the Association of Universities for Research in Astronomy,
Inc. under NASA contract NAS 5-26555. Support for this work was provided by 
NASA through grant number HST-GO-13744.001-A from the Space Telescope Science Institute, 
which is operated by AURA, Inc., under NASA contract NAS 5-26555. I.O. acknowledges
a grant GACR 14-20666P.
    The SDSS is managed by the Astrophysical Research Consortium for the Participating Institutions. 
GALEX is a NASA mission  managed  by  the  Jet  Propulsion  Laboratory.
This research has made
use of the NASA/IPAC Extragalactic Database (NED) which is operated by the
Jet  Propulsion  Laboratory,  California  Institute  of  Technology,  under  contract
with the National Aeronautics and Space Administration.
 \item[Author Contribution] All authors participated at the design of the HST observational
program. Y.I.I. and N.G.G. selected the galaxy sample. T.X.T. and Y.I.I. lead the
observations. G.W. reduced the HST data. I.O. did part of the HST data analysis.
Y.I.I. and D.S. did the SED modelling and interpretation. A.V. and Y.I.I. did the Ly$\alpha$
interpretation. The bulk of the text was written by Y.I.I. 
All authors commented on the manuscript at all stages.
 \item[Competing Interests] Reprints and permissions information is available 
at www.nature.com/reprints. The authors declare no competing financial 
interests.
 \item[Correspondence] Correspondence
should be addressed to Y.I.I.~(email: izotov@mao.kiev.ua).
\end{addendum}

\clearpage
\begin{figure}
\begin{center}
\includegraphics[width=3.45in,angle=0]{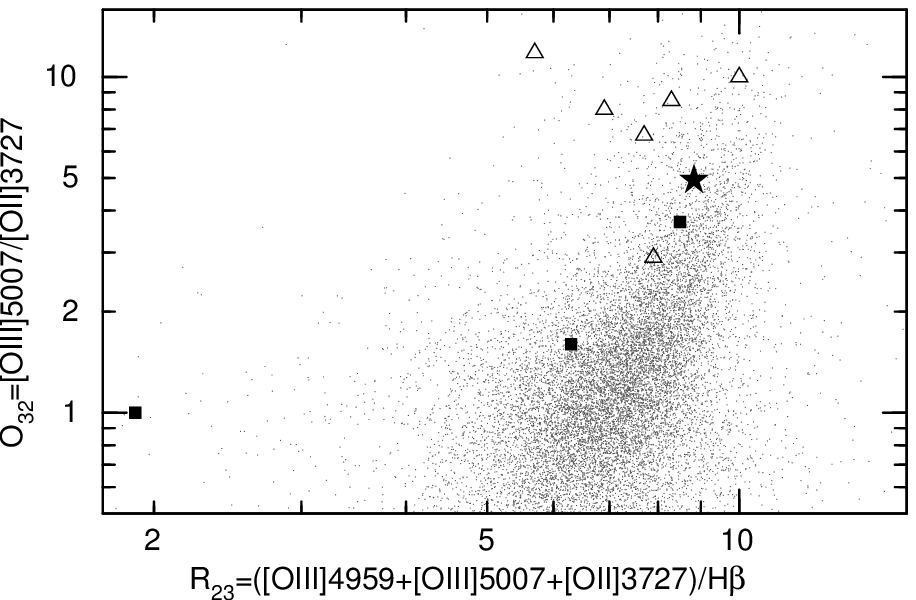}\\
\end{center}

{\bf Figure 1 ${\bf |}$ The O$_{32}$ -- R$_{23}$ diagram for star-forming galaxies.}
The quantity R$_{23}$ is the total flux of the strongest oxygen lines in the optical
spectrum relative to H$\beta$. This quantity is used for easier comparison 
with high-redshift LAEs potentially leaking ionizing radiation\cite{NO14}, shown by open
triangles. At low metallicities 12 + logO/H $<$ 8.3 R$_{23}$ increases with the metallicity.
The location of J0925$+$1403 and known low-redshift LyC 
leaking galaxies\cite{L13,B14} are shown by filled star and filled squares, respectively. 
SDSS star-forming galaxies\cite{I14a} are represented by dots.
\end{figure}

\begin{figure}
\begin{center}
\includegraphics[width=3.45in,angle=0]{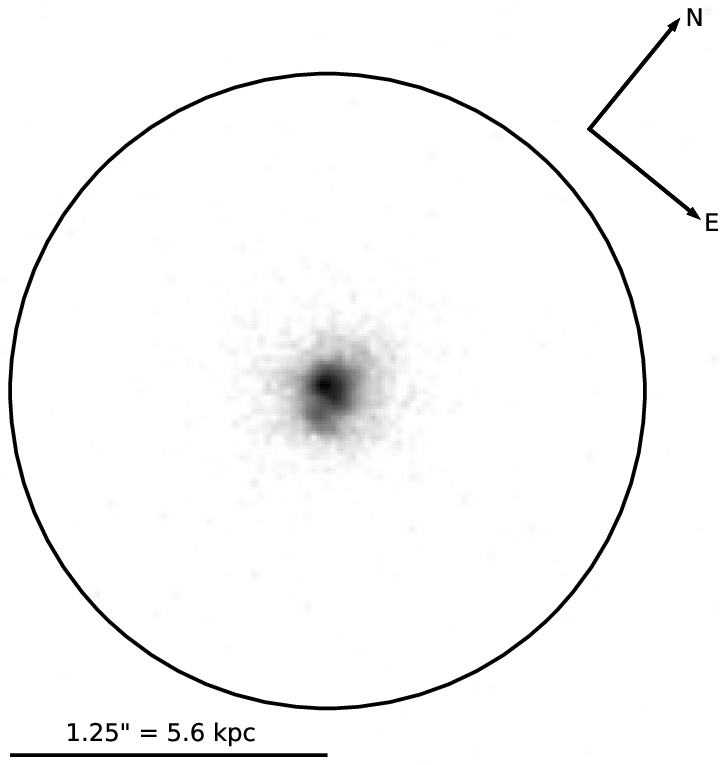}\\
\end{center}
{\bf Figure 2 ${\bf |}$ The Hubble Space Telescope near-ultraviolet
image of J0925+1403.}
The galaxy with a linear diameter $\sim$ 2 kpc consists of
two compact star-forming regions superimposed on an extended low-surface-brightness
component. The spectroscopic aperture with a diameter of 2.5$''$ is shown by a circle.
\end{figure}

\begin{figure}
\begin{center}
\includegraphics[width=3.45in,angle=0]{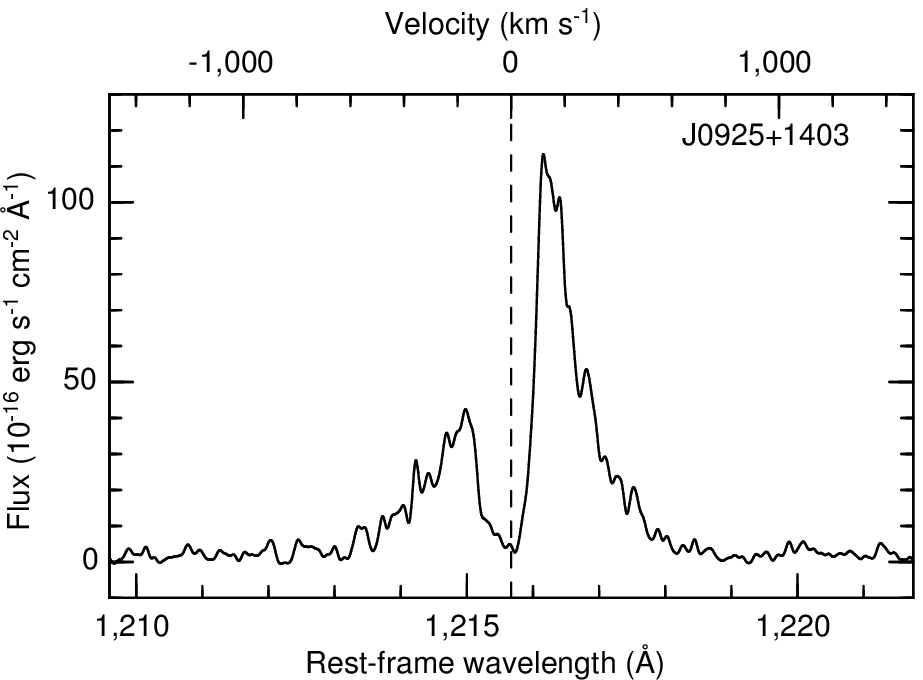}\\
\end{center}

{\bf Figure 3 ${\bf |}$ The double-peaked Ly$\alpha$ emission line in the 
COS spectrum of J0925$+$1403.}
The center of the line is shown by a vertical dashed line.
The small separation of the two emission peaks is indicative of a low
HI column density according to radiation transfer models for spherical geometry\cite{V15}.
\end{figure}

\begin{figure}
\begin{center}
\includegraphics[width=3.45in,angle=-0]{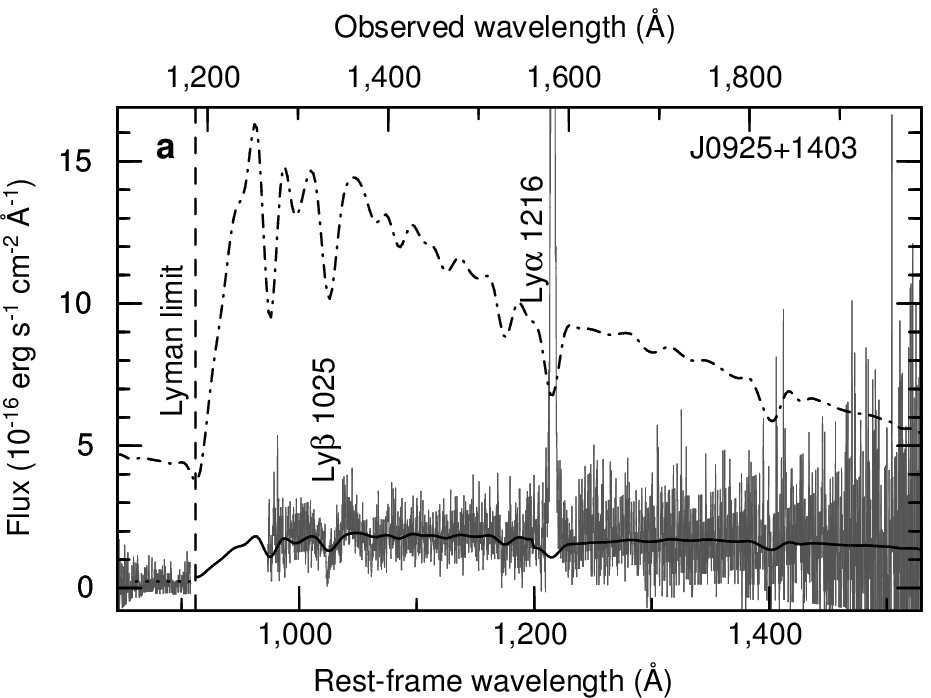}\\
\vspace{0.3cm}
\includegraphics[width=3.45in,angle=-0]{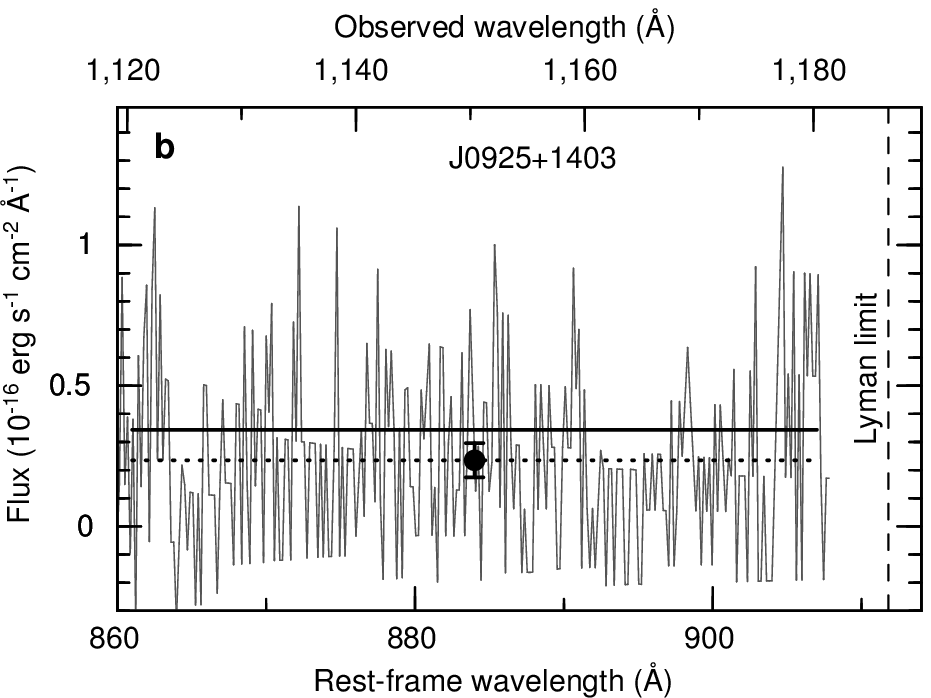}\\
\end{center}

{\bf Figure 4 ${\bf |}$ The COS spectrum of J0925$+$1403.} 
{\bf a,} On top of the spectrum (grey line) is superposed 
the modelled spectral energy distribution (SED), reddened by both 
the Milky Way and internal extinctions (black solid line).  
The unreddened SED is shown by the dash-dotted line.
The dotted line indicates the average observed flux density of the LyC.
{\bf b,} The spectrum in the wavelength range 860 -- 908 \AA. 
The average observed value with the 3$\sigma$ error bar is shown by filled 
circle and the dotted line. The solid line represents the LyC level after 
correction for the Milky Way extinction.
\end{figure}

\clearpage

\begin{methods}

\subsection{The sample of compact star-forming galaxies.}

   A sample of compact star-forming galaxies was selected from the 
spectroscopic data base of the SDSS Data Release 10 (DR10)\cite{I15} by 
applying the selection criteria:
1) $R_{50}$ $\leq$ 3$''$, where $R_{50}$ is the galaxy's Petrosian radius, within 
which 50\% of the galaxy's flux in the SDSS $r$
band is contained; 2) spiral galaxies were excluded; 3) the emission-line ratio 
[OIII]$\lambda$4959/H$\beta$ is $\geq$ 1 to include only galaxies with 
high-excitation HII regions, ensuring an accurate determination of their extinctions 
and chemical compositions; 4) galaxies with AGN activity were excluded 
using line ratios as described below.
Applying these criteria, 5182 galaxies were found in the redshift range 0 $<$ $z$ $<$ 1.
Twenty five galaxies of these have O$_{32}$ $\geq 5$ and are located in
the redshift range 0.30 -- 0.35, making direct LyC observations possible. 
J0925$+$1403  with equatorial coordinates R.A.=09:25:32.37 and Dec=$+$14:03:13.06
was selected for the HST/COS
UV observations as one of the brightest objects in the sample of 25 galaxies,
with a far-ultraviolet magnitude of 20.7 mag from GALEX.
The high rest-frame equivalent width EW$_{{\rm H}\beta}$ = 177\AA\ of the H$\beta$ emission
line in the SDSS spectrum of J0925$+$1403 suggests active ongoing star formation.

The [OIII]$\lambda$5007/H$\beta$ -- 
[NII]$\lambda$6584/H$\alpha$ diagnostic diagram\cite{BPT81} of extragalactic
objects with emission-line spectra, shown in Extended Data Fig. 1, is useful for 
discriminating between star-forming galaxies 
(SFG) and active galactic nuclei (AGN) as ionising sources. The solid line represents the 
model line\cite{K03} that separates the two types of objects. Extended Data Fig. 1 clearly shows that the gas in J0925$+$1403 is ionised by hot stars in
star-forming regions.

\subsection{Custom HST/COS G140L Data Reduction.}

J0925$+$1403 was observed with the HST/COS grating G140L for two orbits
at a central wavelength of 1280\,\AA\ in all four focal-plane offset positions to minimise
fixed-pattern noise and to patch grid-wire shadows and other detector blemishes.
The data were reduced with CALCOS v2.21 and custom software specifically designed
for faint HST/COS targets\cite{W11,S12}, improving upon previous results on
LyC leakage obtained with the default CALCOS pipeline products\cite{B14}.
As our spectra were taken at COS Detector Lifetime Position 3, we did not employ
pulse height filtering to preserve source flux in the presence of detector gain sag from
previous usage of COS at the nearby COS Lifetime Position 1.
We used boxcar extraction in a narrow 25~pixel rectangular window that preserves
spectrophotometry for compact sources (Fig. 2) while minimising the background.
The COS background is dominated by detector dark current, which was subtracted
in post-processing using scaled dark exposures to account for gain sag in the COS aperture.
Specifically, we co-added dark exposures taken in similar orbital conditions within 3 months
of the science observations, smoothed the dark current in the COS aperture with a 500-pixel
running average to remove Poisson fluctuations, and rescaled the smoothed dark current to
the science observations using unilluminated regions of the detector.
The resulting estimate of the COS dark current is sufficiently accurate ($\simeq 5$\,\%) over
the wavelength range of interest (1100--1180\,\AA) to not affect our analysis. Subexposures
were co-added by summing raw counts and the smoothed dark current, accounting for
differing pixel exposure times due to detector blemishes before converting to flux via the
COS calibration curve. This procedure preserves the Poisson counts of faint sources.
Airglow contamination (NI\,$\lambda 1134$\,\AA\ and OI\,$\lambda 1304$\,\AA)
was eliminated by considering only data taken in orbital night in the affected wavelength ranges.
We also verified with orbital night data that scattered HI Ly$\alpha$ airglow is negligible
in the LyC of J0925$+$1403. The sum of diffuse open-shutter backgrounds
(earthshine, zodiacal light, Galactic emission\cite{M14}) is a factor $\simeq 30$ smaller than
the measured flux, leading us to the conclusion that the measured flux is indeed source LyC
flux and not unaccounted background.

\subsection{Extinction in the optical range.}

Using the observed decrement of several
hydrogen Balmer emission lines, we corrected the line fluxes relative to the 
H$\beta$ flux for two effects: (1) reddening adopting
the extinction curve\cite{C89} and 
(2) underlying hydrogen stellar absorption. Both are obtained simultaneously by an
iterative procedure\cite{ITL94}. The quantity derived from the
hydrogen Balmer decrement, is the extinction coefficient $C_{{\rm H}\beta}$,
corresponding to the extinction at the H$\beta$ wavelength 
$A_{{\rm H}\beta} = 2.512 \times C_{{\rm H}\beta}$.
Using the $A_\lambda$/$A_V$ fits\cite{C89}, we approximate
the ratio $C_{{\rm H}\beta}$/$A_V$ by the relation
\begin{equation}
C_{{\rm H}\beta}/A_V = 0.6633-0.1317R_V+0.0294R_V^2-0.0024R_V^3
\end{equation}
with an accuracy better than 0.1\% in the range $R_V$ = $A_V$/$E_{B-V}$ = 2.0 -- 4.0.
We note that $C_{{\rm H}\beta}$/$A_V$ is weakly dependent on $R_V$, decreasing by 
$<$10\% with $R_V$ increasing from 2 to 4.

The correction for reddening was done in two steps. First, the observed 
spectrum, uncorrected for redshift, was corrected for the Milky Way extinction with 
$A_{V,{\rm MW}}$ = 0.084 mag and $R_{V,{\rm MW}}$ = 3.1 (NASA Extragalactic 
Database), corresponding to $C_{{\rm H}\beta,{\rm MW}}$ = 0.039. 
Then, the rest-frame spectrum was corrected for the internal extinction of J0925$+$1403, 
obtained to be $C_{{\rm H}\beta,{\rm int}}$ = 0.175 from the hydrogen Balmer decrement, 
corresponding to $A_{V,{\rm int}}$ = 0.38 mag for $R_{V,{\rm int}}$ = 3.1 and 
$A_{V,{\rm int}}$ = 0.36 mag for $R_{V,{\rm int}}$ = 2.4. The extinction-corrected 
emission-line fluxes relative to the H$\beta$ emission line flux and the rest-frame 
equivalent widths are shown in Extended Data Table 1.

\subsection{Element abundances.}

The [OIII]$\lambda$4363 \AA\ emission line is detected in the SDSS spectrum
of J0925$+$1403. This allows for a reliable oxygen abundance determination 
using the direct $T_{\rm e}$-method. The temperature $T_{\rm e}$(OIII) is 
calculated based on the 
[OIII]$\lambda$4363/($\lambda$4959+$\lambda$5007) 
line ratio\cite{I06}. We adopt a two-zone photoionised HII
region model: a high-ionisation zone with temperature $T_{\rm e}$(OIII), 
where [OIII] lines originate, and a 
low-ionisation zone with temperature $T_{\rm e}$(OII), where [OII]
lines originate. For $T_{\rm e}$(OII), we use
a model relation between the electron temperatures $T_{\rm e}$(OIII) and
$T_{\rm e}$(OII)\cite{I06}. 
Ionic and total abundances of oxygen, nitrogen and neon are derived 
using expressions for ionic abundances and ionisation correction 
factors\cite{I06}. The derived temperatures and element abundances 
are shown in Extended Data Table~2. 
We note that the weak [SII] $\lambda$6717, 6731 emission 
lines, which are used for the electron number density determination,
can not be measured because they are located in the noisy part of the
J0925+1403 spectrum. Therefore, we adopted the value $N_{\rm e}$ = 100 cm$^{-3}$, 
typical of extragalactic HII regions.

\subsection{Fitting of spectral energy distribution and determination of galaxy global 
parameters.}

The luminosity, stellar mass and star-formation rate are important 
global galaxy
characteristics. For the determination of the stellar mass, 
we follow earlier approach for modelling dwarf galaxies\cite{G06,I11,I14a,I14}.
The method is based on fitting a series of model SEDs to the observed one 
and finding the best fit. The fit was performed 
for the SDSS spectrum over the entire
observed spectral range of 3900--9200 \AA.
As the SED is the sum of both stellar and ionised gas emission,
its shape depends on the relative contribution of these two
components. In J0925$+$1403, with a rest-frame H$\beta$ 
equivalent width EW$_{{\rm H}\beta}$ = 177 \AA, the ionised gas continuum is strong 
and should be subtracted before determining the stellar mass.

We carried out a series of Monte Carlo simulations to
reproduce the SED of J0925$+$1403. To derive
the stellar SED, we use a grid of instantaneous burst SEDs 
in a wide range of ages from 0.0 Myr to 15 Gyr, calculated with {\em Starburst99}\cite{L99,L14}. 
We adopted Geneva stellar evolution tracks for nonrotating and rotating stars\cite{M94,E12}
and Padova stellar evolution tracks\cite{G00}. Various models of stellar 
atmospheres were used\cite{L97,S92,HM98,P98}, and we adopt two different
stellar initial mass functions\cite{S55,K01}.
Then the SED with any star-formation 
history can be obtained by integrating the instantaneous burst SEDs over
time with a specified time-varying star-formation rate. 
The SED of the gaseous continuum was taken into account.
It included hydrogen and helium free-bound, free-free, and two-photon emission\cite{A84}.

The star-formation history is approximated assuming a recent 
short burst with age $t_{\rm y}$ $<$ 10 Myr, which accounts 
for the young stellar population, and a prior continuous star formation 
for the older stars during the time interval between $t_i$ and $t_f$
($t_f < t_i$ and zero age is now).
The contribution of each stellar population to the SED was parameterized by the
ratio $b$ = $M_{\rm y}$/$M_{\rm o}$, where $M_{\rm y}$ and $M_{\rm o}$ 
are respectively the masses of the young and old stellar populations.
Then the total stellar mass is $M_*$= $M_{\rm y}$ + $M_{\rm o}$.

We calculated 10$^5$ Monte Carlo models by randomly varying $t_{\rm y}$, $t_i$,
$t_f$, and $b$, while other parameters, such as evolutionary tracks, stellar atmosphere
models, initial mass function and metallicity are kept fixed. In all cases we used
models with metallicities, which best match the metallicity of J0925$+$1403. 
We also calculate the equivalent width EW$_{{\rm H}\beta}$ for each model. 
The best solution is required to fulfill two conditions. First, only models,
in which the modelled equivalent width EW$_{{\rm H}\beta}$
of the H$\beta$ emission line agrees with the observed value within 5 \%, were selected. 
Second, the best modelled SED among selected models for each set of fixed parameters was 
found from $\chi ^2$ minimization of 
the deviation between the modelled and the observed continuum in five wavelength ranges, 
which are free of the emission lines and residuals of the night-sky lines.
We found best solutions for 33 combinations of evolutionary tracks, stellar atmosphere
models and initial mass functions to investigate the dependence of the LyC flux density
on the input parameters. All these solutions provide almost equally good fits 
at rest-frame wavelengths greater 912\AA\ and small variations of LyC.
The optical spectrum of J0925$+$1403 with the overlaid SED of one out of the 33 
best models is shown in Extended Data Fig. 2. Here we adopted 
a Salpeter IMF\cite{S55},
Padova evolutionary tracks\cite{G00} and a combination of stellar atmosphere 
models\cite{L97,S92}.

The modelled SED in the UV range is shown in Fig. 4a by a dash-dotted line. We note that
the amount of the flux decrease at the Lyman break at a fixed starburst age depends on the
adopted stellar evolution tracks, stellar atmosphere models, and initial mass function.
In particular, the Lyman break in models with non-rotating stars is stronger, than
in models with rotating stars.
However, we find from the SED fitting that the intrinsic flux density
$I$($\lambda$912) of the LyC varies only by $\sim$10 \% in the various sets
of models. This is primarily due to the fact that to fit the observed 
EW$_{{\rm H}\beta}$, the starburst age in models with rotating stars should be 
greater than that in models with non-rotating stars\cite{L14}. 
For young bursts with the age of 0 -- 3 Myr we obtained the approximate relation 
$I$($\lambda$912) $\sim$ $I$(H$\beta$)/12, which may be used to estimate the intrinsic
LyC flux density in galaxies with $f_{\rm esc}$ $<<$ 1 without modelling.
We also note that this relation is not much changed if continuous star formation is adopted 
for the young stellar population instead of an instantaneous burst. This is due to the 
dominant contribution of the 0 -- 3 Myr stellar population to both the LyC and H$\beta$ 
fluxes, while the contribution of older stars is small.

To derive global characteristics, the observed fluxes were transformed to
luminosities adopting the luminosity distance 1620 Mpc\cite{W06} and 
the cosmological parameters 
$H_0$=67.1 km s$^{-1}$Mpc$^{-1}$, $\Omega_\Lambda$=0.682, 
$\Omega_m$=0.318\cite{P14}.
Additionally, for the determination of the H$\beta$ luminosity $L_{{\rm H}\beta}$
and star formation rate (SFR), the H$\beta$ flux density corrected for aperture effects
using the relation 2.512$^{r({\rm app})-r}$, where $r$ and $r$(app) are respectively the 
SDSS $r$-band total magnitude and the magnitude within the round spectroscopic 3$''$
diameter aperture. The aperture correction for J0925$+$1403 is 1.54.
The derived H$\beta$ luminosity (Extended Data Table 3) corresponds to
a number of ionising photons $Q_{\rm H}$ = 4.94$\times$10$^{54}$ s$^{-1}$,
in good agreement with the $Q_{\rm H}$ obtained from the integration of
the number of ionising photons in the modelled intrinsic spectrum 
(it is only $\sim$8\% lower). The
corresponding star-formation rate is derived using the standard
Kennicutt relation\cite{K98}.

\subsection{Reddening law for J0925$+$1403.}

The comparison of the modelled SEDs with the observed photometric and spectroscopic data
in the entire UV and optical range 
allows to verify, which extinction curve is most applicable to 
J0925$+$1403. It is known that the extinction curve in the UV range for
the Small Magellanic Cloud with oxygen abundance 12+logO/H~$\sim$~8.1 
is much steeper than the Milky Way extinction
curve, and is characterised by $R_V$ $\sim$ 2.7\cite{B85,GC98,G03}. On the other hand,
the extinction curve in the optical range is insensitive to variations of $R_V$. The
oxygen abundance of J0925$+$1403 is lower, $\sim$7.9. Therefore, we may expect 
the extinction curve in this galaxy to be characterized by an even lower $R_V$.

The observed UV and optical spectra are shown in Extended Data Figure~3 by
grey lines. Their fluxes are consistent with the SDSS and GALEX photometric fluxes shown by
the black symbols. To fit these observational data we reddened the intrinsic modelled 
SED, keeping the same Milky Way extinction $C_{{\rm H}\beta,{\rm MW}}$ = 0.039, 
$R_{V,{\rm MW}}$ = 3.1, and the internal extinction 
$C_{{\rm H}\beta,{\rm int}}$ = 0.175, but varying 
$R_{V,{\rm int}}$ = 3.1 (dotted line), 2.7 (dashed line) and 2.4 (solid line).
The modelled SEDs were calculated with Padova stellar evolution models\cite{G00}, 
a combination of stellar atmosphere models\cite{L97,S92}
and a Salpeter initial mass function\cite{S55}.
It is seen that the SEDs reddened with $R_{V,{\rm int}}$ = 3.1 and 2.7 do not fit the observed
fluxes in the UV range. A higher extinction coefficient $C_{{\rm H}\beta,{\rm int}}$
would be needed in the UV. This is difficult to understand in the framework of a model with
an uniform dust distribution, characterised by a single value of the extinction. 
On the other hand, a SED reddened by the same extinction $C_{{\rm H}\beta,{\rm int}}$ = 0.175, but with $R_{V,{\rm int}}$ = 2.4, nicely 
reproduces the observed fluxes. If we use different stellar atmosphere models for the SED modelling,
we find somewhat larger $R_{V,{\rm int}}$ in some cases, but not exceeding 2.7.

\subsection{Relative Lyman continuum escape fraction.}

To compare with other studies, we also estimate the LyC escape fraction from the 
often-used equation\cite{L13}
\begin{equation}
f_{\rm esc}=f_{\rm esc,rel} \times 10^{-0.4\times A_{1500}} =
\frac{(f_{1500}/f_{900})_{\rm int}}{(f_{1500}/f_{900})_{\rm obs}}\times 
10^{-0.4\times A_{1500}},
\end{equation}
where $f_{900}$ and $f_{1500}$ are the flux densities at rest-frame wavelengths 900\AA\
and 1500\AA, respectively, the subscripts ``int'' and ``obs'' denote mean intrinsic and 
observed flux densities, $(f_{1500}/f_{900})_{\rm int}$ = 1.36 from our SED fits, $(f_{1500}/f_{900})_{\rm obs}$
= 5.79, and $A_{1500}$ = 1.58 mag is the internal extinction at 1500\AA\ (Fig. 4a). 
From this we obtain the relative escape fraction $f_{\rm esc,rel} = 23.6$\%, and the absolute 
$f_{\rm esc} =5.5$\%. The latter value is lower than our above determination for two reasons.
First, Eq. 2 does not take into account  the fact that the expected extinction at 900\AA\ is 
higher than at 1500\AA. And second, the observed  flux density $f_{900}$ must be corrected for 
foreground extinction from the Milky Way. Therefore, $f_{\rm esc}$ obtained with Eq. 2 is 
underestimated. 

\subsection{Code availability.}

We have opted not to make the codes for the custom HST/COS data 
reduction\cite{W11} and the galaxy spectral energy distribution 
fitting\cite{I15} available because they are not yet adapted for
public use.
The {\em Starburst99} code used to generate spectral energy distribution of single stellar
populations is available at  http://www.stsci.edu/science/starburst99/docs/parameters.html.

\end{methods}


\pagestyle{empty}

\setcounter{figure}{0}
\renewcommand{\figurename}{Extended Data Figure 1}

\clearpage

\begin{figure}
\begin{center}
\includegraphics[width=3.45in,angle=-0]{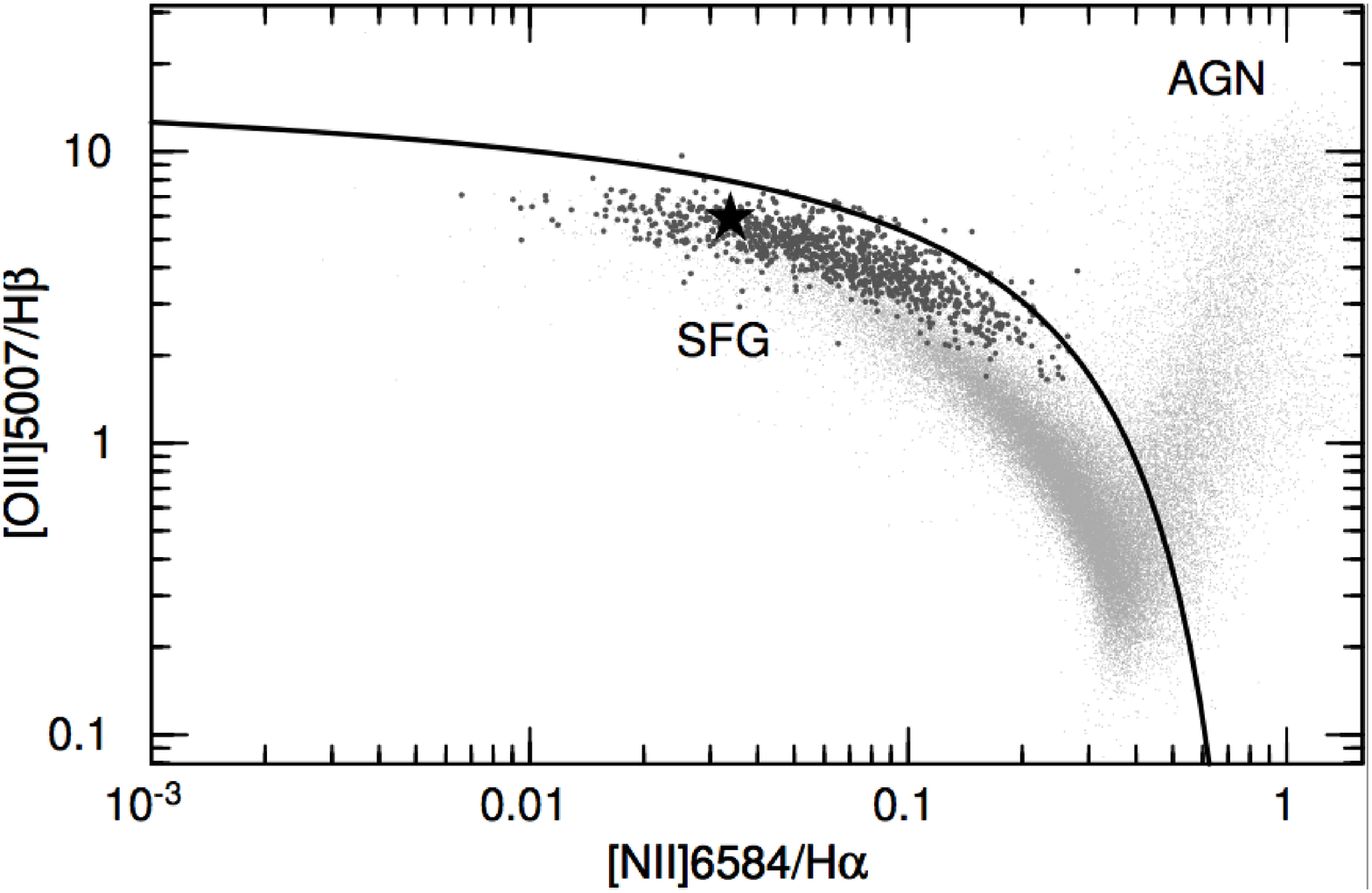}
\end{center}
{\bf Extended Data Figure 1 ${\bf |}$ The diagnostic
diagram\cite{BPT81} for narrow emission lines.}
The galaxy J0925$+$1403 is shown by a large filled star,
and the Luminous Compact Galaxies\cite{I11} by small dark-grey circles.
Also, plotted are the 100,000 emission-line galaxies from SDSS DR7 (cloud
of light-grey dots). The solid line\cite{K03} separates 
star-forming galaxies (SFG) from active galactic nuclei (AGN).
\end{figure}

\setcounter{figure}{0}
\renewcommand{\figurename}{Figure S2}


\begin{figure}
\begin{center}
\includegraphics[width=3.45in,angle=-0]{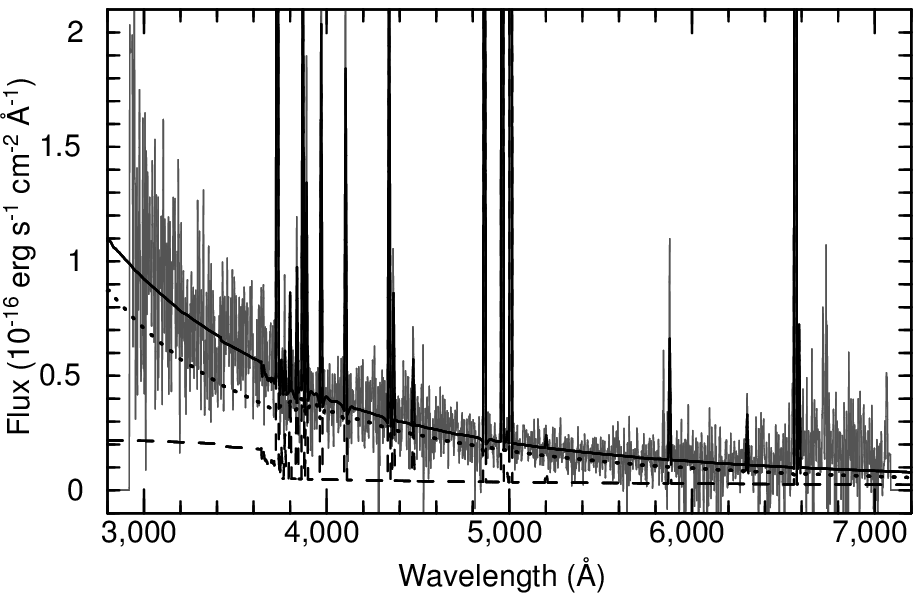}
\end{center}
{\bf Extended Data Figure 2 ${\bf |}$ SED fitting of the optical spectrum of 
J0925$+$1403.}
The rest-frame extinction-corrected spectrum is shown by a grey line.
The stellar, ionised gas, and total modelled SEDs are shown by black dotted, dashed 
and solid lines, respectively.
\end{figure}

\begin{figure}
\begin{center}
\includegraphics[width=3.45in,angle=-0]{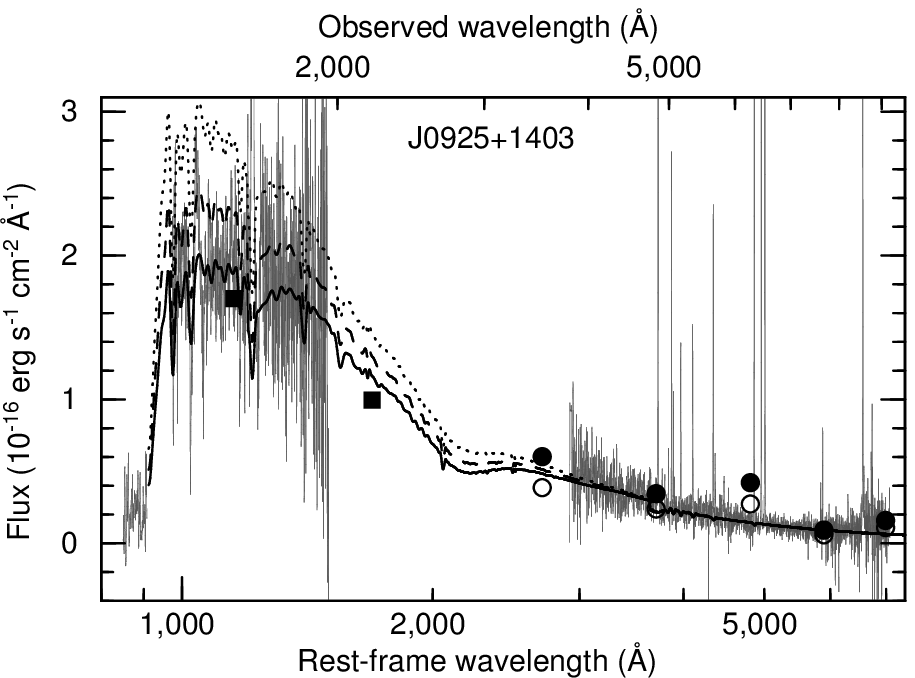}
\end{center}
{\bf Extended Data Figure 3 ${\bf |}$ A comparison of the observed UV and optical
spectrum with the modelled SED.}
The observed spectrum is shown by a grey line. The total GALEX and SDSS
photometric fluxes are represented by filled squares and filled circles, respectively,
while the SDSS photometric fluxes within a round spectroscopic aperture of 3$''$ in 
diameter are shown by open circles.
Modelled SEDs, which are reddened by the Milky Way with $R_{V,{\rm MW}}$ = 3.1 and 
internal extinction with different values of $R_{V,{\rm int}}$, are shown by black lines. 
Dotted, dashed and solid lines correspond to $R_{V,{\rm int}}$ = 3.1, 2.7, and 2.4,
respectively. 
\end{figure}

\pagestyle{empty}

\setcounter{table}{0}
\renewcommand{\tablename}{Table S1}

\begin{table}
\centering
{\bf Extended Data Table 1 ${\bf |}$ Emission-line fluxes and equivalent 
widths in the optical spectrum.}\\ 
\begin{tabular}{lcrr} \\ \hline
Line&Wavelength&100$\times$$I$($\lambda$)/$I$(H$\beta$)$^{\rm \dag}$&EW($\lambda$) \\
    &    (\AA) &   & (\AA)   \\ \hline
$[$OII$]$          &3727&125.5$\pm$ 4.6&  77 \\
H9                &3835&  9.1$\pm$ 2.0&   7 \\
$[$NeIII$]$        &3868& 47.5$\pm$ 1.9&  33 \\
HeI+H8            &3889& 17.1$\pm$ 2.3&  12 \\
$[$NeIII$]$+H7     &3968& 32.0$\pm$ 2.6&  22 \\
H$\delta$          &4101& 28.7$\pm$ 2.3&  22 \\
H$\gamma$       &4340& 43.7$\pm$ 2.5&  36 \\
$[$OIII$]$         &4363& 11.7$\pm$ 0.6&  10 \\
HeI                &4471&  6.1$\pm$ 0.5&   6 \\
H$\beta$          &4861&100.0$\pm$ 3.6& 177 \\
$[$OIII$]$         &4959&199.9$\pm$ 6.7& 306 \\
$[$OIII$]$         &5007&608.1$\pm$ 20.&1174 \\
HeI                &5876& 11.3$\pm$ 0.7&  22 \\
$[$OI$]$           &6300&  4.6$\pm$ 0.5&  11 \\
H$\alpha$         &6563&280.2$\pm$ 9.9& 732 \\
$[$NII$]$          &6584& 13.2$\pm$ 0.8&  26 \\
\hline \\
\end{tabular}

Footnote:~~~~~~~~~~~~~~~~~~~~~~~~~~~~~~~~~~~~~~~~~~~~~~~~~~~~~~~~~~~~~~~~~~~~~~~~~~~~~

$^{\rm \dag}$Extinction-corrected flux relative to the extinction-corrected

~~~~~~~flux $I$(H$\beta$) = 4.92$\times$10$^{-15}$ erg s$^{-1}$ cm$^{-2}$ of the H$\beta$
emission line, 

multiplied by 100.~~~~~~~~~~~~~~~~~~~~~~~~~~~~~~~~~~~~~~~~~~~~~~~~~~~~~~~~~~~~~~~~
\end{table}

\pagestyle{empty}

\setcounter{table}{0}
\renewcommand{\tablename}{Table S2}

\begin{table}
\centering
{\bf Extended Data Table 2 ${\bf |}$ Physical conditions and chemical composition.} \\ 
\begin{tabular}{lc} \\ \hline
Parameter&Value \\ \hline
$T_{\rm e}$ ($[$OIII$]$), K        & 15010$\pm$410 \\
$T_{\rm e}$ ($[$OII$]$), K        & 14010$\pm$360 \\
$N_{\rm e}$ ($[$SII$]$), cm$^{-2}$ &       100$^{\rm \dag}$ \\ \\
O$^+$/H$^+$$\times$10$^{5}$                &1.42$\pm$0.11 \\
O$^{2+}$/H$^+$$\times$10$^{5}$              &6.65$\pm$0.50 \\
O/H$\times$10$^{5}$                        &8.06$\pm$0.52 \\
12+log O/H                               &7.91$\pm$0.03 \\ \\
N$^+$/H$^+$$\times$10$^{6}$                &1.12$\pm$0.07 \\
ICF(N)$^{\rm \ddag}$                           &5.42 \\
N/H$\times$10$^{6}$                        &6.05$\pm$0.45 \\
log N/O                                  &$-$1.12$\pm$0.04 \\ \\
Ne$^{2+}$/H$^+$$\times$10$^{5}$             &1.27$\pm$0.11 \\
ICF(Ne)$^{\rm \ddag}$                          &1.08 \\
Ne/H$\times$10$^{5}$                       &1.37$\pm$0.13 \\
log Ne/O                                 &$-$0.77$\pm$0.05 \\
\hline \\
\end{tabular}

Footnote:~~~~~~~~~~~~~~~~~~~~~~~~~~~~~~~~~~~~

$^{\rm \dag}$Assumed value.~~~~~~~~~~~~~~~~~~~~~~~~~~~

$^{\rm \ddag}$Ionisation correction factor.~~~~~~~~~
\end{table}

\begin{table}
\centering
{\bf Extended Data Table 3 ${\bf |}$ Global characteristics of J0925+1403.} \\ 
\begin{tabular}{lc} \\ \hline
Parameter&Value \\ \hline
${I_{{\rm H}\beta}}^{\rm \dag}$      &49.2$\pm$1.3 \\
Redshift                           &0.301323\\
Luminosity distance$^{\rm \ddag}$&1620 \\
${L_{{\rm H}\beta}}^{\rm \dag\dag}$      &(2.32$\pm$0.04)$\times$10$^{42}$ \\
SFR$^{\rm \ddag\ddag}$                & 52.2 \\
${Q_{\rm H}}^{\rm *}$           &4.94$\times$10$^{54}$ \\
$Q_{\rm H}$(esc)$^{\rm *}$    &3.86$\times$10$^{53}$ \\
$t$(burst)$^{\rm **}$                &2.6$\pm$0.2 \\
$M_{\rm y}$/$M_\odot$             &(2.4$\pm$0.3)$\times$10$^8$ \\
$M_{\rm *}$/$M_\odot$             &(8.2$\pm$0.7)$\times$10$^8$ \\ 
\hline \\
\end{tabular}

Footnote:~~~~~~~~~~~~~~~~~~~~~~~~~~~~~~~~~~~~~~~~~~~~~~~~~~~~~~~~~

\noindent ~~~~~~~~~~~~~$^{\rm \dag}$Extinction-corrected flux density in 10$^{-16}$ erg s$^{-1}$cm$^{-2}$.\\
\noindent $^{\rm \ddag}$in Mpc.~~~~~~~~~~~~~~~~~~~~~~~~~~~~~~~~~~~~~~~~~~~~~~~~~~~~~~~~~~~~~~\\
~~~~~~~~~~~~~~~$^{\rm \dag\dag}$Extinction- and aperture-corrected luminosity in erg s$^{-1}$.\\
~~~~~~~~~~~~~~~~~~~~~~~~~~~$^{\rm \ddag\ddag}$Star-formation rate in $M_\odot$yr$^{-1}$ derived from the H$\beta$ 
luminosity\cite{K98}. \\
~~~~~~~~~~~~~~~~~~~~~~~~~~~~$^{\rm *}$$Q_{\rm H}$ and $Q_{\rm H}$(esc) are the numbers of LyC photons in s$^{-1}$ 
emitted\\
~~~~~~~~~~~~~~~~~~~~~~~~~~~~~~~by massive stars and escaped from the HII region, respectively.\\
$^{\rm **}$Burst age in Myr.~~~~~~~~~~~~~~~~~~~~~~~~~~~~~~~~~~~~~~~~~~~~~
\end{table}

\end{document}